\documentclass[]{emulateapj}
\def\lesssim{\mathrel{\mathpalette\fun <}}
\def\gtrsim{\mathrel{\mathpalette\fun >}}
\def\fun#1#2{\lower3.6pt\vbox{\baselineskip0pt\lineskip.9pt
  \ialign{$\mathsurround=0pt#1\hfil##\hfil$\crcr#2\crcr\sim\crcr}}}
\def\be{\begin{equation}}
\def\ee{\end{equation}}
\def\bea{\begin{eqnarray}}
\def\eea{\end{eqnarray}}

\shorttitle{Fermi acceleration at relativistic shocks}
\shortauthors{Lemoine, Pelletier \& Revenu}

\begin{document}

\title{On the efficiency of Fermi acceleration at relativistic shocks}
\author{Martin Lemoine\altaffilmark{1}}
\author{Guy Pelletier\altaffilmark{2}}
\author{Beno\^\i t Revenu\altaffilmark{3}}
\altaffiltext{1}{Institut d'Astrophysique de Paris,
UMR 7095 CNRS, Universit\'e Pierre \& Marie Curie, 
98 bis boulevard Arago, F-75014 Paris, France; email: lemoine@iap.fr}
\altaffiltext{2}{Laboratoire d'Astrophysique de Grenoble LAOG,
CNRS, Universit\'e Joseph Fourier,
BP~53, F-38041 Grenoble, France; email: guy.pelletier@obs.ujf-grenoble.fr}
\altaffiltext{3}{APC -- Coll\`ege de France,
11 place Marcelin Berthelot, F-75231 Paris Cedex 05, France,
\& SUBATECH, BP20722, F-44307 Nantes Cedex 03, France; email: benoit.revenu@in2p3.fr}

\begin{abstract}
It is shown that Fermi acceleration at an ultra-relativistic shock
wave cannot operate on a particle for more than 1 {\small 1/2} Fermi
cycle (i.e., $\rm u\rightarrow d\rightarrow u\rightarrow d$) if the
particle Larmor radius is much smaller than the coherence length of
the magnetic field on both sides of the shock, as is usually assumed.
This conclusion is shown to be in excellent agreement with recent
numerical simulations. We thus argue that efficient Fermi acceleration
at ultra-relativistic shock waves requires significant non-linear
processing of the far upstream magnetic field with strong
amplification of the small scale magnetic power. The streaming or
transverse Weibel instabilities are likely to play a key r\^ole in
this respect.
\end{abstract}

\section{Introduction}
Fermi acceleration of charged particles bouncing back and forth a
collisionless shock wave is at the heart of a variety of phenomena in
high energy astrophysics. According to standard lore, this includes
the acceleration of electrons at gamma-ray bursts internal/external
relativistic shock waves, whose synchrotron light is interpreted as
the prompt/afterglow radiation.

However the inner workings of Fermi acceleration at relativistic shock
waves remain the subject of intense study and debate, even in the test
particle limit. It has been argued that a universal energy spectral
index $s\simeq 2.2-2.3$ should be expected (e.g., Bednarz \& Ostrowski
1998, Achterberg {\it et al.} 2001, Lemoine \& Pelletier 2003, Ellison
\& Double 2004, Keshet \& Waxman 2005), which would agree nicely with
the index that is inferred from gamma-ray bursts observations. Yet
recent numerical simulations that include more realistic shock
crossing conditions have indicated otherwise (e.g., Lemoine \& Revenu
2006, Niemiec \& Ostrowski 2006), so that the situation is presently
rather confuse. Unfortunately, numerical Monte-Carlo simulations of
Fermi acceleration, although they remain a powerful tool, do not shed
light on the physical mechanisms at work.

In the present {\em Letter}, we offer a new analytical discussion of
Fermi acceleration in the ultra-relativistic regime. We rely on the
observation that, under standard assumptions, the Larmor radius
$r_{\rm L}$ of freshly injected particles is much smaller than the
coherence length $l_{\rm coh}$ of the upstream magnetic field, and we
integrate the equations of motion to first order in the quantity
$r_{\rm L}/l_{\rm coh}$ (Section 2). We thus find that a particle
cannot execute more than 1 1/2 $\rm u\rightarrow d\rightarrow u$
cycles through the shock before escaping downstream, hence Fermi
acceleration is mostly inoperative. This result is related to the very
short return timescale in the ultra-relativistic regime: over its
trajectory, the particle experiences a nearly coherent and mostly
transverse magnetic field, hence the process is akin to superluminal
acceleration in regular magnetic fields discussed by Begelman \& Kirk
(1990). We also show that our analytical predictions agree very well
with numerical simulations. Finally, we argue that the (expected)
non-linear processing of the magnetic field at ultra-relativistic
shock waves and, in particular, the strong amplification of
small-scale power, may be the agent of efficient Fermi acceleration
(Section 3); we suggest new avenues of research in this direction.

\section{Analytical trajectories}

\subsection{Field line curvature}

The upstream magnetic field consists of a regular component $B_0$ and
a turbulent component $\delta B$: $\vec B = \vec B_0 + \vec{\delta
B}$, with $\langle B^2\rangle = B_0^2 + \langle \delta B^2\rangle$.
The turbulence is defined in the wavenumber range $k_{\rm
min}<k<k_{\rm max}$ by its power spectrum $S(k)\propto k^{-\alpha}$,
which is normalized according to: $ \int {\rm d}^3k\,
S(k)\,=\,\langle\delta B^2\rangle$. Downstream and upstream magnetic
fields are related to each other by the MHD shock jump conditions (see
e.g., Kirk \& Duffy 1999). In the case of $\gamma-$ray bursts external
shocks, the inferred fraction of energy density stored in magnetic
turbulence is of the order of a percent (Gruzinov \& Waxman 1999) so
that the magnetic field can be considered as passive. In this limit,
one finds: $B_{\parallel \vert\rm d}\,=\,B_{\parallel\vert\rm u}\
,\quad \vec B_{\perp\vert\rm d}\,=\,R_{\rm sh}\vec B_{\perp\vert\rm
u}$; $B_{\parallel}\equiv\vec B\cdot\vec z$ is the component of the
magnetic along the shock normal $\vec z$, $\vec B_{\perp}$ is the
projection of $\vec B$ on the shock front plane $(\vec x,\vec y)$, and
subscripts $_{\vert\rm d}$ (resp. $_{\vert\rm u}$) indicate that the
quantity is measured in the downstream (resp. upstream) plasma rest
frame. The quantity $R_{\rm sh}\,\equiv\, \Gamma_{\rm sh\vert
u}\beta_{\rm sh\vert u}/(\Gamma_{\rm sh\vert d}\beta_{\rm sh\vert d})$
is the proper shock compression ratio, expressed in terms of
$\beta_{\rm sh\vert u}$ (resp. $\beta_{\rm sh\vert d}$) the shock
velocity measured in the upstream (resp. downstream) rest frame and the
corresponding Lorentz factor $\Gamma_{\rm sh\vert u}$ (resp.
$\Gamma_{\rm sh\vert d}$). In the ultra-relativistic limit
($\Gamma_{\rm sh\vert u}\gg1$): $R_{\rm sh}\simeq \Gamma_{\rm sh\vert
u}\sqrt{8}\gg1 $.

Hence, to an error $\sim{\cal O}(1/\Gamma_{\rm sh\vert u})$ on the
direction of $\vec B$, it is a good approximation to consider that the
magnetic field lies in the transverse $(\vec x,\vec y)$ plane
downstream of the shock. The magnetic field at a given point $\vec
r_0$ on the shock surface can be written in both downstream and
upstream reference frames as follows: \bea \vec B_{\vert\rm d}(\vec
r_0)& \,\simeq\, &B_{\perp\vert\rm d}\cos(\phi_B)\vec x +
B_{\perp\vert\rm d}\sin(\phi_B)\vec y\ ,\nonumber\\ \vec B_{\vert\rm
u}(\vec r_0)& \,=\, &B_{\perp\vert\rm u}\cos(\phi_B)\vec x +
B_{\perp\vert\rm u}\sin(\phi_B)\vec y + B_{\parallel\vert\rm u}\vec z\
.  \eea {\em The phase $\phi_B$ is invariant under the Lorentz
transformation from downstream to upstream}; this observation plays a
key r\^ole in the discussion that follows.

We assume for the time being that the Larmor radius $r_{\rm L}$ of the
test particle is much smaller than the coherence length of the
magnetic field $l_{\rm coh}$; if $\alpha > 3$, $l_{\rm coh}\sim
1/k_{\rm min}$ as the magnetic power is distributed on the largest
spatial scales. Then, since the typical $\rm u\rightarrow d
\rightarrow u$ cycle time through the shock is of order ${\cal
O}\left(r_{\rm L}/\Gamma_{\rm sh\vert u}\right)\ll l_{\rm coh}$
(Achterberg et al. 2001, Lemoine \& Pelletier 2003, Lemoine \& Revenu
2006), {\em in a first approximation one can neglect the magnetic
field line curvature over the trajectory of the particle}.

This approximation may be justified as follows. Consider a particle
moving over a length scale $l\sim r_{\rm L}/\Gamma_{\rm sh\vert u}\ll
l_{\rm coh}$. The radius of curvature ${\cal R}_{>l}$ of the magnetic
field on scales larger than $l$ can be calculated as: \be {\cal
R}_{>l}^{-1} \,\equiv\,\left\langle\left\vert{(\vec
B\cdot\vec\nabla)\vec B\over
B^2}\right\vert^2\right\rangle_{>l}^{1/2}\ .  \ee Assuming that
$\alpha>3$ (or $l_{\rm coh}\sim 1/k_{\rm min}$) and decomposing
$\vec{\delta B}$ in a Fourier series, one finds that indeed, $l/{\cal
R}_{>l}\sim (\delta B/B)^{-1}\,(l/l_{\rm coh})^{(\alpha-3)/2} \ll
1$. Hence the large scale component is approximately uniform over a
length scale $l$. It is safe to neglect the magnetic power on scales
smaller than $l$ since $\delta B_{<l}^2 \sim \delta B^2 (l/l_{\rm
coh})^{\alpha-3}\ll \delta B^2$, the latter being comparable to the
large scale component. If $\alpha<3$, similar conclusions apply, since
the assumption $r_{\rm L}\ll l_{\rm coh}$ translates into $r_{\rm
L}\ll k_{\rm max}^{-1}$, which means that the particle only
experiences a smooth large scale magnetic field.

Now, if the magnetic field is approximately regular over the path of
the particle in a $\rm u\rightarrow d\rightarrow u$ cycle, Fermi
acceleration in the ultra-relativistic regime becomes similar to
superluminal acceleration in a fully regular magnetic field, which is
known to be inefficient (Begelman \& Kirk 1990). In the following, we
extend the discussion of these authors and compare the predictions to
numerical simulations of particle propagation in realistic turbulence.

\subsection{Analytical trajectories}

\noindent{\it Upstream.} The equation of motion reads: \be {{\rm
d}\vec\beta\over{\rm d}t}\,=\,\Omega_{\rm L}\,{\vec\beta\times\vec
B\over B}\ , \ee with $\vec\beta$ the velocity of the particle and
$\Omega_{\rm L}=c/r_{\rm L}$ the Larmor frequency. Shock crossing from
downstream toward upstream requires $\beta_{\parallel,\,\rm i}^{\rm
(u)}\geq\beta_{\rm sh\vert u}$ with $\beta_{\parallel,\,\rm i}^{\rm
(u)}$ the ingress component of the velocity along the shock
normal. Hence $\beta_{\perp,\,\rm i}^{\rm (u)}\sim {\cal
O}(1/\Gamma_{\rm sh\vert u})$. By working to first order in
$1/\Gamma_{\rm sh\vert u}$, Achterberg {\it et al.} (2001) were able to
obtain analytically the particle trajectory and its direction at shock
recrossing $\rm u\rightarrow d$. 
Assuming that $\phi_B=0$ so that the transverse component of $\vec B$
lies along $\vec x$, one obtains the outgoing velocity vector as: \bea
\beta^{({\rm u})}_{x,\,\rm f}&\,\simeq\,& \beta^{({\rm u})}_{x,\,\rm
i}\ ,\nonumber\\ \beta^{({\rm u})}_{y,\,\rm f}&\,\simeq\,& -{1\over
2}\beta^{({\rm u})}_{y,\,\rm i} + \left[{3\over \Gamma_{\rm sh\vert
u}^2} - 3\beta^{({\rm u})\,2}_{x,\,\rm i} - {3\over 4}\beta^{({\rm
u})\,2}_{y,\,\rm i}\right]^{1/2} \ ,\nonumber\\ \beta^{({\rm
u})\,2}_{z,\,\rm f}&\,=\,&1-\beta^{({\rm u})\,2}_{x,\,\rm
f}-\beta^{({\rm u})\,2}_{y,\,\rm f}\ .
\label{eq:u2d}\eea

\noindent{\it Downstream.} There we must proceed differently as the
return timescale $\sim {\cal O}(r_{\rm L}/c)$ can no longer be treated
as a small quantity ($r_{\rm L}$ is evaluated in the downstream rest
frame; Lemoine \& Revenu 2006).

We may assume that $\phi_B=0$ since the phase is preserved by the
Lorentz transformation; furthermore shock compression results in a
magnetic field essentially oriented transversally to the shock
normal. To this order of approximation, the trajectory along the shock
normal reads: \be \Omega_{\rm L}z(t)\,=\, \beta^{({\rm
d})}_{\perp,\,\rm i}\sin(\phi_{\rm i})\left[\cos(\Omega_{\rm L}
t)-1\right] + \beta^{({\rm d})}_{\parallel,\,\rm i}\sin(\Omega_{\rm L}
t)\ , \ee where $\phi_{\rm i}$ denotes the phase of the ingress
velocity vector in the $(x,y)$ plane: $\beta^{({\rm d})}_{x,\,\rm
i}\equiv\beta^{({\rm d})}_{\perp,\,\rm i}\cos(\phi_{\rm i})$,
$\beta^{({\rm d})}_{y,\,\rm i}\equiv\beta^{({\rm d})}_{\perp,\,\rm
i}\sin(\phi_{\rm i})$.  The shock front follows the trajectory:
$z_{\rm sh\vert d}( t)=\beta_{\rm sh\vert d} t$ and return to the
shock will occur if and when: \be \sin(\phi_{\rm i})\,=\,g(\hat t
)\,\equiv\,{\beta_{\rm sh\vert d}\hat t - \beta_{\parallel,\,\rm
i}\sin(\hat t)\over \beta^{({\rm d})}_{\perp,\,\rm i}\left[\cos(\hat
t)-1\right]}\ ,
\label{eq:return_cond}\ee with $\hat t=\Omega_{\rm L}t$.  The function
$g(\hat t)$ diverges toward $-\infty$ for $\hat t\rightarrow 0, 2\pi$
and its derivative is monotonous in the interval $\hat t\in ]0,2\pi[$.
Hence return to the shock can occur if and only if the maximum of
$g(\hat t)$ exceeds the value $\sin(\phi_{\rm i})$. Note also that
$g(\hat t)$ is always negative since by assumption $\beta^{({\rm
d})}_{\parallel,\,\rm i}\leq \beta_{\rm sh\vert d}$. Therefore {\em a
necessary condition for return to the shock front is $\phi_{\rm i} \in
[-\pi,0]$, or equivalently $\beta^{({\rm d})}_{y,\,\rm i}\leq0$}.

Once the time $t_{\rm d}$ of shock return has been determined
(numerically), the outgoing velocity vector can be derived from the
solutions to the equations of motion: \bea \beta^{({\rm d})}_{x,\,\rm
f}&\,\simeq\,&\beta^{({\rm d})}_{x,\,\rm i}\ ,\nonumber\\ \beta^{({\rm
d})}_{y,\,\rm f}&\,\simeq\,&\beta^{({\rm d})}_{y,\,\rm
i}\cos(\Omega_{\rm L} t_{\rm d}) + \beta^{({\rm d})}_{z,\,\rm
i}\sin(\Omega_{\rm L} t_{\rm d})\, \nonumber\\ \beta^{({\rm
d})}_{z,\,\rm f}&\,\simeq\,&\beta^{({\rm d})}_{z,\,\rm
i}\cos(\Omega_{\rm L} t_{\rm d}) - \beta^{({\rm d})}_{y,\,\rm
i}\sin(\Omega_{\rm L} t_{\rm d})\ .\label{eq:d2u} \eea

\subsection{Mappings: downstream to upstream and vice-versa}
Equations~(\ref{eq:u2d}) and (\ref{eq:d2u}) define mappings from the
ingress to the egress angles on either side of the shock. The ingress
angles in one rest frame are related to the egress angles in the other
rest frame by the Lorentz transformations: \bea \phi^{({\rm u})}_{\rm
i}&\,=\,&\phi^{({\rm d})}_{\rm f},\quad\phi^{({\rm d})}_{\rm
i}\,=\,\phi^{({\rm u})}_{\rm f}\ ,\nonumber\\ \beta^{({\rm
u})}_{\parallel,\,\rm i}&\,=\,&{\beta^{({\rm d})}_{\parallel,\,\rm
f}+\beta_{\rm rel}\over 1 + \beta^{({\rm d})}_{\parallel,\,\rm
f}\beta_{\rm rel}},\quad \beta^{({\rm d})}_{\parallel,\,\rm
i}\,=\,{\beta^{({\rm u})}_{\parallel,\,\rm f}-\beta_{\rm rel}\over 1 -
\beta^{({\rm u})}_{\parallel,\,\rm f}\beta_{\rm rel}}\ , \eea where
$\beta_{\rm rel}\equiv (\beta_{\rm sh\vert u}-\beta_{\rm sh\vert
d})/(1- \beta_{\rm sh\vert u}\beta_{\rm sh\vert d})$ is the relative
velocity between the upstream and downstream rest frames. Using these
mappings and transformations, one can follow the trajectory of a
particle. Since the cycle time is of order $r_{\rm L}/\Gamma_{\rm
sh}\ll l_{\rm coh}$, it is reasonable to assume that $\phi_B=0$
remains constant from one Fermi cycle to the next.

Now, in Section 2.2.2, we argued that $\beta^{({\rm d})}_{y,\,\rm
i}\leq0$ is a necessary condition for the particle to be able to
return to the shock. However, as the particle travels upstream and
exits back toward downstream, its outgoing velocity is given by
Eq.~(\ref{eq:u2d}) and it can be shown that $\beta^{({\rm
u})}_{y,\,\rm f}\geq0$ irrespectively of the upstream ingress
angle. In effect, for a given $\beta^{({\rm u})}_{y,\,\rm i}$, the
quantity $\beta^{({\rm u})}_{y,\,\rm f}$ is minimal when $\beta^{({\rm
u})}_{x,\,\rm i}$ is maximal, i.e. when $\beta^{({\rm u})\,2}_{x,\,\rm
i}=1-\beta_{\rm sh\vert u}^2-\beta^{({\rm u})\,2}_{y,\,\rm i}$. Then
the final $\beta^{({\rm u})}_{y,\,\rm f}=\left(-\beta^{({\rm
u})}_{y,\,\rm i} + 3\vert\beta^{({\rm u})}_{y,\,\rm i}\vert\right)/2
\geq 0$. The minimum is then 0, which corresponds to a particle
entering upstream along $\vec x$ (tangentially to the shock surface),
i.e. $\beta^{({\rm u})}_{z,\,\rm i}=\beta_{\rm sh}$, $\beta^{({\rm
u})}_{y,\,\rm i}=0$.  {\em Hence, if a particle that travels
downstream is able once to return to the shock, it will not do so in
the subsequent cycle}.

\begin{figure}[ht]
  \centering\includegraphics[width=0.45\textwidth,clip=true]{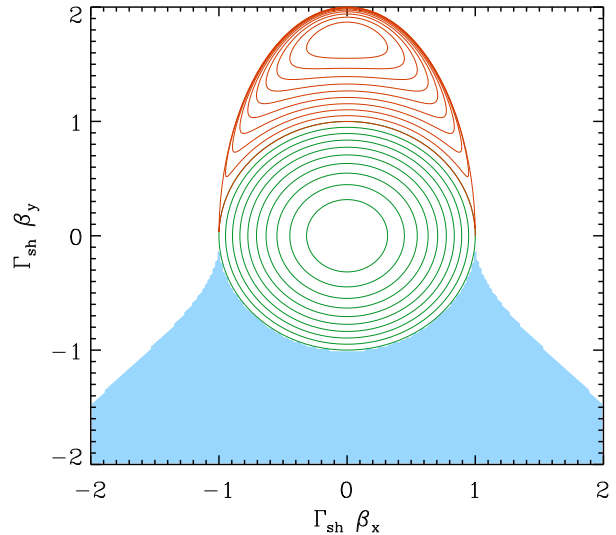}
  \caption[...]{ Mapping from downstream to upstream and back to
upstream as measured in the upstream rest frame, in the plane
transverse to the shock front; $\beta_x$ and $\beta_y$ are the
velocity components in this plane (note the enhancement by
$\Gamma_{\rm sh\vert u}$ on each axis). The solid blue area shows the
region of egress upstream coordinates which permits the particle to
return to the shock from downstream. The green circled area shows how
the original downstream particle population maps upon entering
upstream, and the red kidney-shaped region shows the mapping of this
population on exit from upstream.
\label{fig:f1}
}
\end{figure}

A quantitative assessment of this discussion is shown in
Fig.~\ref{fig:f1} which presents the locii of ingress and egress
velocity vectors in the $(x,y)$ plane as seen in the upstream rest
frame. The blue area shows the region of egress $\beta^{({\rm
u})}_{x,\,\rm f}$ and $\beta^{({\rm u})}_{y,\,\rm f}$ (equivalently
ingress as seen from downstream) for which the particle is bound to
return to the shock. The green circles show the ingress $\beta^{({\rm
u})}_{x,\,\rm i}$ and $\beta^{({\rm u})}_{y,\,\rm i}$ of a particle
that crosses toward upstream. The various circles correspond to
different values of $\beta^{({\rm u})}_{z,\,\rm i}$ upon entry; the
radii of these circles are bounded by the shock crossing condition
$\beta^{({\rm u})}_{z,\,\rm i}\geq\beta_{\rm sh\vert u}$. Finally, the
red kidney shaped forms map these ingress upstream velocities into the
egress velocities, according to Eq.~(\ref{eq:u2d}). The fact that
these kidney shaped forms do not overlap anywhere with the blue area
confirms that at most one and a half cycle $\rm u\rightarrow
d\rightarrow u\rightarrow d$ is permitted.

\subsection{Comparison with numerical work}

\begin{figure}[ht]
  \centering\includegraphics[width=0.45\textwidth,clip=true]{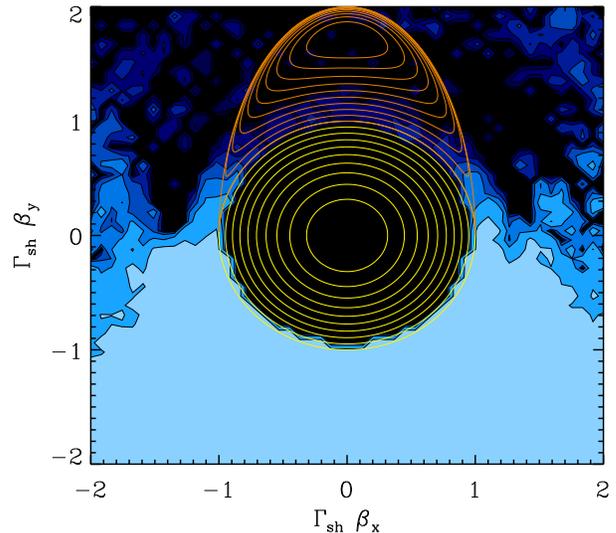}
  \caption[...]{ Same as Fig.~\ref{fig:f1} but using results from the
numerical simulation of the trajectories of $\sim10^6$ particles in
shock compressed turbulence (see text for details). Each contour
represents a drop in return probability by a factor 2.4; the lighter
the shading, the higher the return probability, black corresponding to
zero. The yellow circles and red kidney-shaped curves are the same as
those shown in Fig.~\ref{fig:f1}. }
\label{fig:f2}
\end{figure}

The previous discussion relies on several approximations, most notably
that the field lines can be considered as straight over the trajectory
of the particle. Comparison of the previous results with numerical
simulations of particle propagation in refined descriptions of the
magnetic field are best suited to assess the error that results from
these approximations. Figure~\ref{fig:f2}, which shows the contour
plot of the return probability defined as a function of
$\beta_{x,\,\rm i}$ and $\beta_{y,\,\rm i}$, can be directly compared
to the blue area of Fig.~\ref{fig:f1}. Indeed, the agreement is
excellent. The parameters of the simulations whose results are shown
in Fig.~\ref{fig:f2} are as follows: $r_{\rm L}/L_{\rm max}= 7\times10^{-4}$,
$\alpha=11/3$ (Kolmogorov turbulence), $B_0=0$ (pure turbulence) and
$\Gamma_{\rm sh\vert u}=38$. The numerical procedure used to follow
the particle trajectory has been described in Lemoine \& Pelletier
(2003) and Lemoine \& Revenu (2006).

This discussion also explains the results of recent Monte-Carlo
simulations. For instance, Niemiec \& Ostrowski (2006) report that
Fermi acceleration is inefficient in the ultra-relativistic regime for
upstream Kolmogorov turbulence; their simulations indicate very steep
spectra if any, in good agreement with the present discussion. In
contrast, other studies of Fermi acceleration obtain powerlaw spectra
of various spectral indices. However, one can check that these latter
studies have, one way or another, either assumed an isotropic
downstream turbulence, or implicitly marginalized over the angle
between the particle trajectory and $\vec B_\perp$ at shock crossing,
which amounts to picking $\phi_B$ at random in each half-cycle. In the
light of the above discussion, it is then easy to understand why Fermi
acceleration seemed efficient in these studies.

Strictly speaking, our results do not apply to scale invariant
turbulence, i.e. $\alpha=3$. However, results of numerical simulations
for this particular case are similar to those shown in
Fig.~\ref{fig:f2}; the return probability is non-zero everywhere but
it is ten times lower in the kidney-shaped region than in the negative
$\beta_{y,{\rm i}}$ area. This suggests that quite steep powerlaw
spectra should emerge from Fermi acceleration in such turbulence.

\section{Discussion}
Fermi acceleration is thus inefficient at ultra-relativistic shock
waves if the Larmor radius $r_{\rm L}$ of injected particles is much
smaller than the coherence length $l_{\rm coh}$ of the turbulent
magnetic field on both sides of the shock. This does not mean that
Fermi acceleration is bound to fail. In particular, if $r_{\rm L}\gg
l_{\rm coh}$, powerlaw spectra must emerge as the memory of the
magnetic field direction (in the transverse plane) at shock crossing
is erased during propagation in small scale turbulence (as we have
checked numerically). The final value of the spectral
index will depend on the transport properties of the particle in this
small-scale turbulence. It is difficult to probe this regime using
numerical simulations as integration timescales become large (the
scattering timescale $\gg r_{\rm L}/c$).  On analytical grounds, one
expects $s\simeq 2.3$ if the small scale turbulence is isotropic
downstream of the shock wave (Keshet \& Waxman 2005); if it is
anisotropic, one ought to expect a different value however.

Although Fermi acceleration could operate efficiently on high energy
initial seed particles with $r_{\rm L}\gg l_{\rm coh}$, the abundance
of such particles is generally so low in realistic astrophysical shock
wave environments that the injection efficiency would be extremely
small (see Gallant \& Achterberg 1999 for instance). 

In the above context, the interpretation of the afterglow emission of
$\gamma-$ray bursts as the synchrotron radiation of electrons
accelerated at the ultra-relativistic external shocks ($\Gamma_{\rm
sh\vert u}\ga 100$) becomes particularly enlightening. The success of
this model indeed requires both efficient Fermi acceleration as well
as very significant amplification of the interstellar magnetic field
(Gruzinov \& Waxman 1999).  Hence it is tempting to tie these two
facts together and to wonder whether this non-linear MHD processing
could not be the agent of efficient Fermi acceleration.

One proposal discussed so far is the transverse Weibel instability
which could possibly produce sufficiently strong magnetic fields on
very small spatial scales \\ $\sim 10^5\,{\rm cm}\,(\Gamma_{\rm sh\vert
u}/10)^{-1/2}(n_e/1\,{\rm cm}^{-3})^{-1/2}$ (Medvedev \& Loeb 1999);
there is however ongoing debate on the lifetime and strength at
saturation of the magnetic field (e.g., Wiersma \& Achterberg 2004;
Lyubarsky \& Eichler 2006). Nonetheless, such a small-scale turbulence
should result in powerlaw spectra of accelerated particles; albeit
the value of the resulting spectral index is not known.

Recently, it has been suggested that the generalization of the
streaming instability to the relativistic regime could amplify the
magnetic field to the values required by $\gamma-$ray bursts
observations (Milosavljevic \& Nakar 2006).  Due to the very short
upstream return timescale $\sim r_{\rm L}/\Gamma_{\rm sh\vert u}$, the
particle can never stream too far ahead of the shock so that the
turbulence is generated on small scales $\sim 10^7-10^8\,{\rm cm}\,\ll
r_{\rm L}$ (Milosavljevic \& Nakar 2006). Therefore, one naturally
expects in this case too that Fermi acceleration would be efficient,
here as well, one needs to understand the turbulence properties before
conclusions can be drawn on the index $s$.

To summarize, we have shown in Section 2 that Fermi acceleration
cannot operate successfully at ultra-relativistic shock waves if one
assumes (somewhat na\"\i vely) large-scale turbulence on both sides of
the shock wave. The conclusions of the present discussion are thus
more optimistic and open a wealth of new possibilities; in particular
they suggest that the success of Fermi acceleration is intimately
connected with the mechanism of magnetic field amplification in the
shock vicinity. The comprehension of Fermi acceleration will
eventually require understanding the generation of the magnetic field,
deriving the properties of the turbulence as well as characterizing
the transport of accelerated particles in this possibly anisotropic
turbulence.\smallskip

\noindent{\bf Note added:} while this work was being completed, a
recent preprint by Niemiec {\it et al.} (2006) appeared, reporting on
Fermi acceleration with small-scale turbulence. Although their
simulations are limited to $\Gamma_{\rm sh\vert u}= 10$, these authors
observe that the inclusion of small scale turbulence allows powerlaw
spectra to emerge through Fermi acceleration, in good agreement with
the above discussion.

\end{document}